\documentclass[final]{agujournal2019} 
\usepackage{url} 
\usepackage{lineno}
\usepackage[inline]{trackchanges} 
\usepackage{soul}
\usepackage{graphicx,color}
\usepackage{tikz}
\usepackage{amsthm,amsmath}
\usepackage[utf8]{inputenc}  
\usepackage{wasysym}
\usepackage[authoryear,round]{natbib}
\let\cite\citep
\usepackage{chemfig}
\usepackage{lineno}
\usepackage{here}
\usepackage{xurl}

\usepackage{amsmath,amssymb}
\usepackage{mathtools}
\usepackage{bm}
\usepackage{lastpage}
\usepackage{ascmac}
\usepackage{physics2}
\usephysicsmodule{ab}
\usephysicsmodule{ab.braket}
\usephysicsmodule{xmat}
\usepackage{fixdif}
\usepackage{derivative}
\usepackage{mathrsfs}
\usepackage{tensor}
\usepackage[
  separate-uncertainty=true,
  range-phrase={\text{ $\sim$ }}, 
  per-mode=symbol
]{siunitx}
\usepackage{array}
\usepackage[export]{adjustbox}
\usepackage{booktabs}

\newcommand{\scalebar}[6][white]{
 \begin{tikzpicture}
  \node[anchor=south west,inner sep=0] (image) { #2 };
  \begin{scope}[x={(image.south east)},y={(image.north west)}]
   \draw [#1, line width=0.2em] (0.04,1.2em) -- node[below,inner sep=0.1em, font=\footnotesize] {\qty{#5}{#6}} (#5*#4/#3+0.04,1.2em);
  \end{scope}
 \end{tikzpicture}
}

\graphicspath{{img/}}

\draftfalse

\journalname{}

\begin{document}

\title{Scaling of microcraters with molten rims derived from laser-induced cratering experiments}

\authors{
                  Soki Sato\affil{1}, 
                  Hitoshi Shibayama\affil{2}, 
                  Kaori Nagai\affil{2}, 
                  Hiroaki Katsuragi\affil{1}
	      }
     
\affiliation{1}{Department of Earth and Space Science, The University of Osaka,  1-1, Machikaneyama, Toyonaka, Osaka, 560-0043, JAPAN}  
\affiliation{2}{College of Industrial Technology, Nihon University, 1-2-1, Izumicho, Narashino city, Chiba, 275-8575, JAPAN}

\begin{abstract}
Microcraters with molten rims are widely observed in returned samples from the Moon and asteroids. These features reflect the conditions of small-scale impact events, but the impact velocities required for their formation are not well understood.
In this study, we use well-controlled laser irradiation on rock surfaces as an experimental analog to investigate the formation of these molten-rim craters. 
Specifically, we derive a scaling relation between crater volume, total irradiation energy, and energy loss associated with thermal diffusion. The experimental results indicate that molten-rim structures of microcraters can be formed in the range of $10^2$–$10^3$~m/s impact velocity. This value is consistent with some previous studies which suggested that the origin of microcraters is secondary impact events. These findings provide experimental constraints on the formation mechanisms of molten-rim microcraters and offer a new perspective on impact processes recorded on planetary surfaces. The estimated result is also consistent with the impact velocity required to produce the observed wavy rim protrusions due to a Rayleigh-Taylor hydrodynamic instability ($\sim 10^2$~m/s).
\end{abstract}

\section{Introduction}
Most of solid bodies in the Solar System, including the Moon and Mars, are covered with a carpet of impact craters. The majority of these craters are thought to originate from meteoritic impacts. Although the Earth is also subjected to meteorite bombardment, its atmosphere shields the surface from most small impactors. Therefore, it is difficult to find small-scale craters on the Earth. Furthermore, intense weathering and erosion, together with active tectonic processes, erase all the impact records at all scales over relatively short geological timescales.  In contrast, the Moon lacks such modification processes. As a result, an enormous number of craters have been preserved on the lunar surface.  Small bodies such as asteroids are also covered with craters. However, because their environmental conditions, particularly gravity, differ significantly from those of larger bodies such as the Moon, crater morphologies exhibit notable differences. On small bodies such as asteroids, for example, meteorite impacts can readily excite global seismic vibrations affecting the entire body. Repeated exposure to such global shaking can efficiently relax crater topography~\cite{RICHARDSONJR:2005}. 

While typical craters are commonly envisioned as large bowl-shaped depressions, craters also exist on microscopic scales. Here, the microscopic scale refers to sizes around the order of micrometers.  The discovery of microcraters has been achieved mainly through the detailed studies of returned samples.  Lunar microcraters were extensively studied in the 1970s~\cite{Brownlee:1973,Schneider:1973,Fechtig:1974,Horz:1975,Flavill:1978}. Some meteorites also possess microcraters~\cite{Brownlee:1973s,Goswami:1976}. Although relatively large ($\sim 10$~$\mu$m) microcraters have a characteristic spall zone, smaller ($\lesssim 1$~$\mu$m) ones show a pit structure with a pronounced rim of molten target material~\cite{Horz:1975}. However, the physical mechanism responsible for these morphological variations has long remained unclear.

Microcraters have also been identified recently in returned samples from asteroids and Moon. Careful observations of particles returned from asteroids confirmed the presence of microcraters on their surfaces~\cite{Nakamura:2012,Matsumoto:2018,Matsumoto:2024}. Similar sub-micro-size craters were discovered recently on the surface of Chang'e-5 lunar soils as well~\cite{Gu:2025}.  Such microcraters are considered to be formed by secondary impacts~\cite{Harries:2016,Matsumoto:2018,Gu:2025}. The rims of these microcraters exhibit distinctive morphologies characterized by multiple protruding wavy structures. These features resemble the so-called “milk crown” phenomenon, a crown-shaped structure formed when a droplet impacts a liquid surface. Splashing phenomena caused by fluid impacts have been studied since the nineteenth century~\cite{Worthington:1897}. If the rims were molten by the impact energy, hydrodynamic instability might cause the wavy structure resulting in milk-crown-like feature~\cite{Katsuragi:2015}. 

Despite these observations, the specific range of impact velocities capable of producing molten rim structures remains poorly constrained. 
Altogether, the central open question addressed in this study is: under what impact conditions do molten and wavy rim structures form on microcraters? However, direct observation alone cannot constrain these conditions, so experimental approaches are needed.

The fundamental physical processes involved in crater formation and degradation have long been investigated~\cite{Melosh:1988,Melosh:2011}.
However, a unified theoretical framework capable of consistently describing the entire sequence of crater formation has not yet been fully established. At present, analyses of crater formation require the understanding of multiple complex physical processes. Consequently, the intricate details of cratering processes have not been fully understood. In particular, the formation of microcraters on the order of micrometers in size involves physical processes distinct from large-scale cratering. For instance, the dominance of surface tension over gravity governs the formation. In addition, the rapid melt solidification preserves micro-scale cratering. These small-scale cratering processes have been poorly understood compared to large-scale cratering processes. 
Although melt is also generated during large-scale crater formation, the behavior of large volumes of melt is dominated by gravity, and surface-tension effects become negligible in large scale. Therefore, the direct observation of capillary-related features can be regarded as a distinctive characteristic unique to microcraters. This point is key to understanding the formation of molten wavy rim structures.

\begin{figure}
   \centering
    \includegraphics[width=0.3\linewidth]{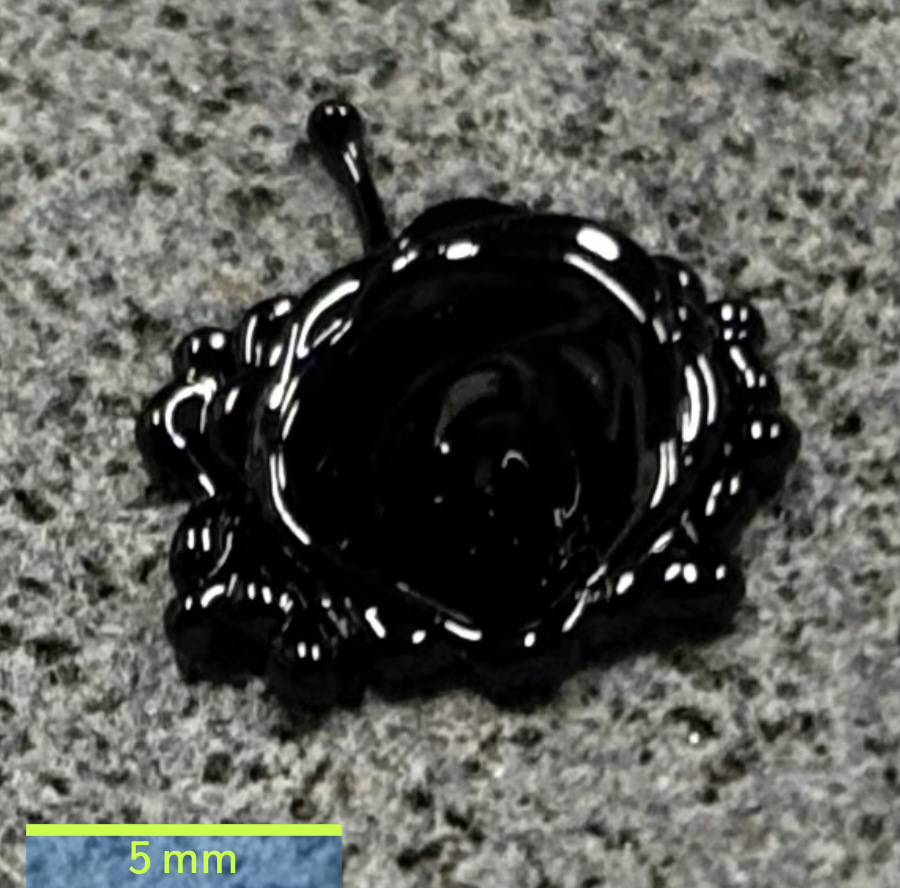}
    \caption{An example of crater with molten rim structure. This crater was formed by a test laser irradiation.}
    \label{fig:example}
\end{figure}

A useful clue to understanding these complex behaviors comes from the field of stone fabrication for architectural engineering. 
Fundamental studies of laser-based material cutting have been extensively conducted~\cite{Nagai:2018,Nagai:2021,Nagai:2022}. From an engineering perspective, a key objective is to achieve cutting with high energy efficiency. In addition, in laser surface processing, it is important to form holes or cuts without generating cracks or excessive ejection of molten material. Under such conditions, however, it has been reported that molten material can sometimes form glass-like residues during laser processing~\cite{Nagai:2022}. A typical example of such crater shapes is shown in Fig.~\ref{fig:example}. The produced structure is formed by the material melting and subsequent rapid cooling.

To investigate the physical conditions of molten rim formation, we utilize laser-induced cratering as an experimental analog. During laser irradiation, a portion of the incident energy is consumed by rock vaporization, while the remainder contributes to heating and melting the target, leading to the formation of molten rims. Crucially, the recoil pressure generated by this rapid vaporization acts as a mechanical driver for crater excavation, which is physically analogous to the shock compression in hypervelocity impacts. In our experiments, the typical irradiation duration is $\sim$10~s and the resultant crater size is $\sim$10~mm.
Because these timescale and spatial scale are quite different between microcraters and laser cutting, we carefully consider a scaling relation which is non-dimensionalized.  From the scaling relation obtained in this study, we discuss the impact conditions to form molten wavy structures on microcrater rims. Note that this study is quite different from the microscopic laser irradiation experiments which directly mimicking explosion-like energy release~\cite{Guskov:2001,Craciun:2002,Marston:2019}. 

The objective of this study is to establish a scaling relation for laser-induced crater formation and to apply the obtained scaling to the estimate of formation conditions of molten-rim microcraters observed on planetary returned samples. Although the temporal scale of laser irradiation is much longer than that of hypervelocity impacts, the relevant physical processes can be compared by only using dimensionless parameters that relate energy deposition to thermal diffusion length.

\section{Materials and methods}
\label{sec:experiment}

\subsection{Experimental setup and samples}
Laser irradiation was applied to the sample surface to induce melting and evaporation of the material. During irradiation, the surface behavior of the sample was recorded using a high-speed camera (Photron, INFINICAM).  The frame rate was set to 2000~fps, with an image resolution of 1246 $\times$ 496~pixels, corresponding to a spatial resolution of 0.103~mm/pixel. To prevent image saturation, ND (neutral density) filters were placed in front of the lens. Depending on the irradiation condition, one or more filters were combined to adjust the transmitted light intensity. 
A fiber laser (wavelength:~1.06~$\mu$m, Nihon University) was used as the laser light source. The average laser power was controlled in the range of $100$-$1000$~W. 

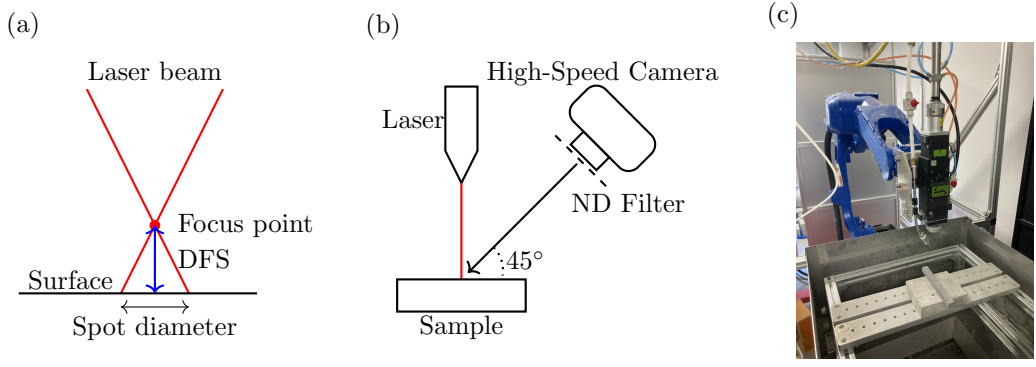
\begin{figure}
    \centering
    \begin{minipage}{0.31\linewidth}
        \centering
        \vbox to 0pt{\vspace{-0.5cm}\hbox{(a)}\vss} 
        \begin{tikzpicture}[scale=0.9]
            \draw[red, thick] (-1, 2) -- (0.5, -1); \draw[red, thick] (1, 2) -- (-0.5, -1);
            \filldraw[red] (0, 0) circle (2pt); \draw[thick] (-2, -1) -- (1.5, -1);
            \draw[<->, blue, thick] (0, -1) -- (0, 0); \draw[<->] (-0.5, -1.2) -- (0.5, -1.2);
            \node[above] at (0, 2) {Laser beam}; \node[right] at (0.2, 0) {Focus point};
            \node[right] at (-2, -0.8) {Surface}; 
            \node[right] at (0.2, -0.5) {DFS};
            \node[below] at (0, -1.2) {Spot diameter};
        \end{tikzpicture}
    \end{minipage}
    \hfill
    \begin{minipage}{0.35\linewidth}
        \centering
        \vbox to 0pt{\vspace{-0.5cm}\hbox{(b)}\vss}
        \begin{tikzpicture}[scale=0.85]
                \draw[thick] (-0.25, 3) -- (-0.25, 2) -- (0, 1.5) -- (0.25, 2) -- (0.25, 3) -- cycle;
                \node at (-0.75, 2.5) {Laser};
                \draw[red, thick] (0, 1.5) -- (0, 0);
                \draw[thick] (-1, 0) rectangle (1, -0.5);
                \node at (0, -0.75) {Sample};
                \draw[thick, rounded corners=5] (2.1, 3.1) -- (3.1, 2.1) -- (2.6, 1.6) -- (1.6, 2.6) -- cycle;
                \draw[thick] (1.9, 2.3) -- (2.3, 1.9) -- (2.1, 1.7) -- (1.7, 2.1) -- cycle;
                \node at (2.2, 3.2) {High-Speed Camera};
                \draw[thick, dashed] (2.25, 1.45) -- (1.45, 2.25);
                \node at (2.6, 1.2) {ND Filter};
                \draw[thick, ->] (1.8, 1.8) -- (0.1, 0.1);
                \draw[thick, dotted] (0.5, 0.5) arc (45:-10:0.5);
        \node at (1, 0.3) {45$^{\circ}$};
        \end{tikzpicture}
    \end{minipage}
    \hfill
    \begin{minipage}{0.28\linewidth}
        \centering
        \vbox to 0pt{\vspace{-0.5cm}\hbox{(c)}\vss}
        \includegraphics[width=0.8\linewidth]{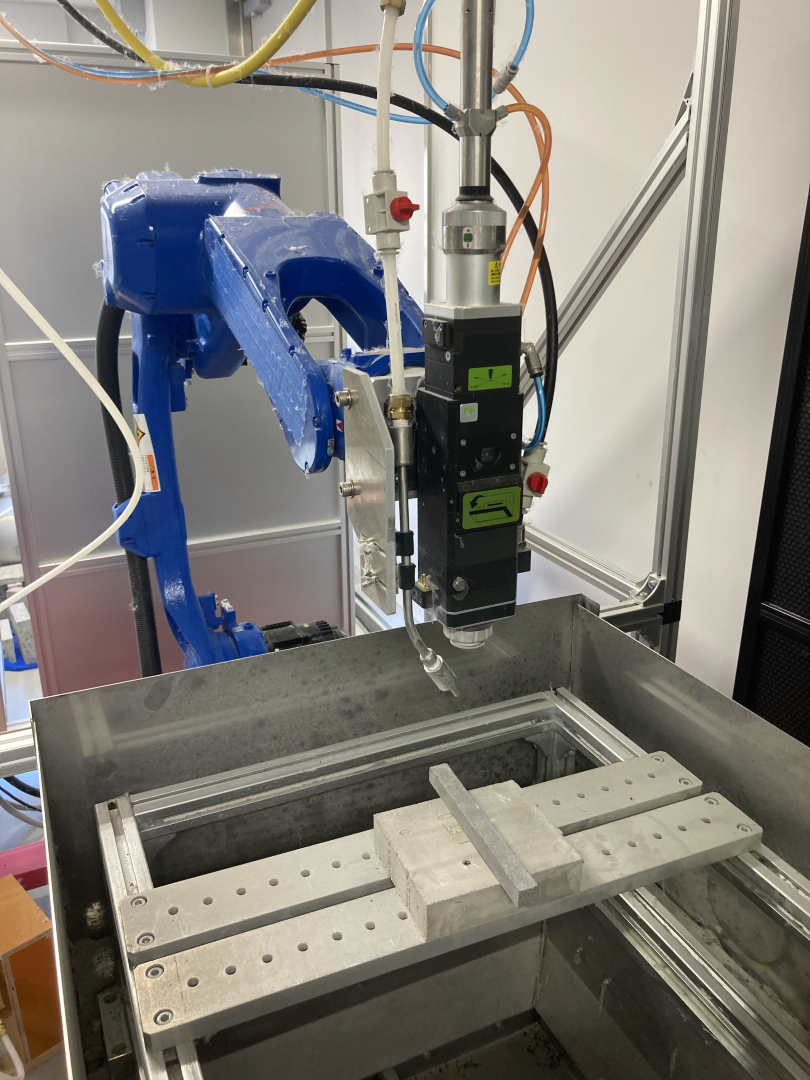}
    \end{minipage}

    \caption{Experimental setup. (a)~Definition of DFS and spot diameter, (b)~Geometrical setup of a laser light source, sample, and high-speed camera. (c)~Actual photo of the experimental setup.}
    \label{fig:laser-setup}
\end{figure}

Andesite samples were used for the irradiation targets as representative igneous rock samples used also in the field of architectural engineering. Each sample was prepared as a block with approximate dimensions of 100 $\times$ 100 $\times$ 60~mm. 
The distance between the laser focal point and the sample surface is referred to as the defocus distance (DFS). The diameter of the laser beam on the sample surface is defined as the spot diameter (Fig.~\ref{fig:laser-setup}(a)).

The laser beam was irradiated perpendicular to the sample surface. A high-speed camera was positioned at an angle of 45$^{\circ}$ from the horizontal plane, and ND filters were attached to the lens to control light intensity depending on the irradiation conditions. A schematic diagram and photograph of the experimental configuration are shown in Fig.~\ref{fig:laser-setup}(b,c).

\subsection{Irradiation Conditions}
\label{subsec:laser-irradiation-conditions}
The average power $P_{\mathrm{ave}}$ of pulsed laser was varied within the range $\sim$100-1000~W. The peak power $P_{\mathrm{peak}}$ was set to $\sim$200 - 2000~W, the pulse duration $t_{\mathrm{p}}$ was 20~ms, and the repetition frequency $f$ was 25~Hz. Thus, the average power was defined as $P_{\mathrm{ave}} = P_{\mathrm{peak}} t_{\mathrm{p}} f$. Cooling-air was supplied at a pressure 0.1~MPa. 
The DFS was varied in 0.1, 0.2, or 0.3~m. The irradiation duration was varied in 5, 10, 25, 50, or 100~s. 
The list of experimental conditions is provided in Table~\ref{tab:all-laser-conditions}.

\begin{table}
    \centering
    \begin{tabular}{@{}lll@{}}
    \hline
         Intensity (W) & DFS (m)   & Irradiation time (s)                                                                        \\
    \hline
         1000 & 0.1 & 5, 10, 25, 50, 100  \\
                      & 0.2  & 5, 10, 25, 50, 100  \\
                      & 0.3  & 5, 10, 25, 50, 100  \\
         750  & 0.1  & 5                                                                           \\
        500  & 0.1  & 5, 10, 25, 50, 100   \\
                     & 0.2  & 5, 10, 25, 50, 100  \\
                     & 0.3  & 5, 10, 25, 50, 100  \\
        350  & 0.1  & 5                                                                           \\
        250  & 0.1  & 5, 10, 25, 50, 100 \\
                     & 0.2  & 5, 100                                                      \\
        150  & 0.1  & 5                                                                           \\
        100  & 0.1  & 5, 10, 25, 50, 100 \\
        \bottomrule
    \hline
    \end{tabular}
    \caption{Experimental (laser irradiation) conditions}
    \label{tab:all-laser-conditions}
\end{table}

\subsection{Crater shape measurement}
\label{subsec:surface-profiling}

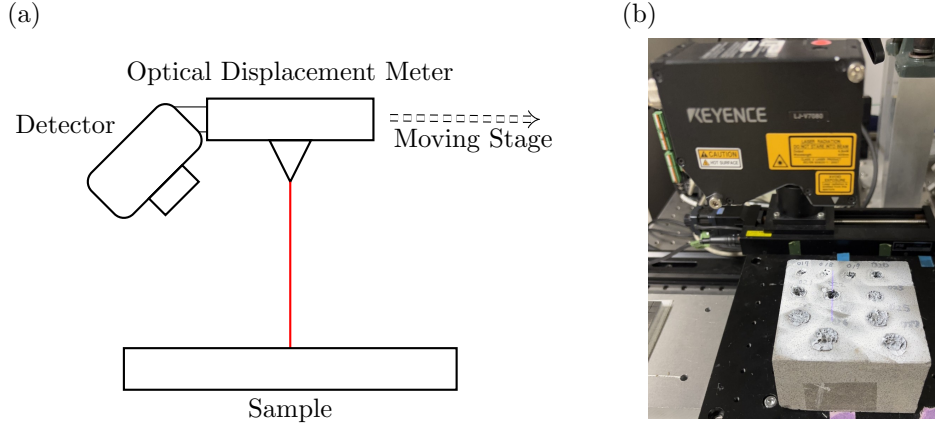
\begin{figure}
    \begin{tabular}{cc}
        \begin{minipage}{0.5\hsize}
            \centering
            \begin{tikzpicture}[scale=1.1]
                \draw[thick] (-1, 3) -- (-1, 2.5)  -- (1, 2.5) -- (1, 3) -- cycle;
                \draw[thick] (-0.25, 2.5) -- (0, 2) -- (0.25, 2.5) -- cycle;
                \node at (0, 3.3) {Optical Displacement Meter};
                \node at (-3.2, 4) {(a)};
                \draw[] (-1, 2.9) -- (-1.4, 2.9);
                \draw[] (-1, 2.6) -- (-1.1, 2.6);
                \draw[thick, rounded corners=5] (-1, 2.5) -- (-2, 1.5) -- (-2.5, 2) -- (-1.5, 3) -- cycle;
                \draw[thick] (-1.3, 2.2) -- (-1.7, 1.8) -- (-1.5, 1.6) -- (-1.1, 2.0) -- cycle;
                \node at (-2.7, 2.7) {Detector};
                \node at (4.2, 4) {(b)};
                \draw[-{Implies}, densely dashed, double distance=2 ] (1.2, 2.77) -- (3, 2.75);
                \node at (2.2, 2.5) {Moving Stage};
                \draw[red, thick] (0, 2) -- (0, 0);
                \draw[thick] (-2, 0) rectangle (2, -0.5);
                \node at (0, -0.75) {Sample};
            \end{tikzpicture}
            \label{fig:odm-schematic}
        \end{minipage} &
        \begin{minipage}{0.45\hsize}
            \centering
            \includegraphics[width=0.6\linewidth]{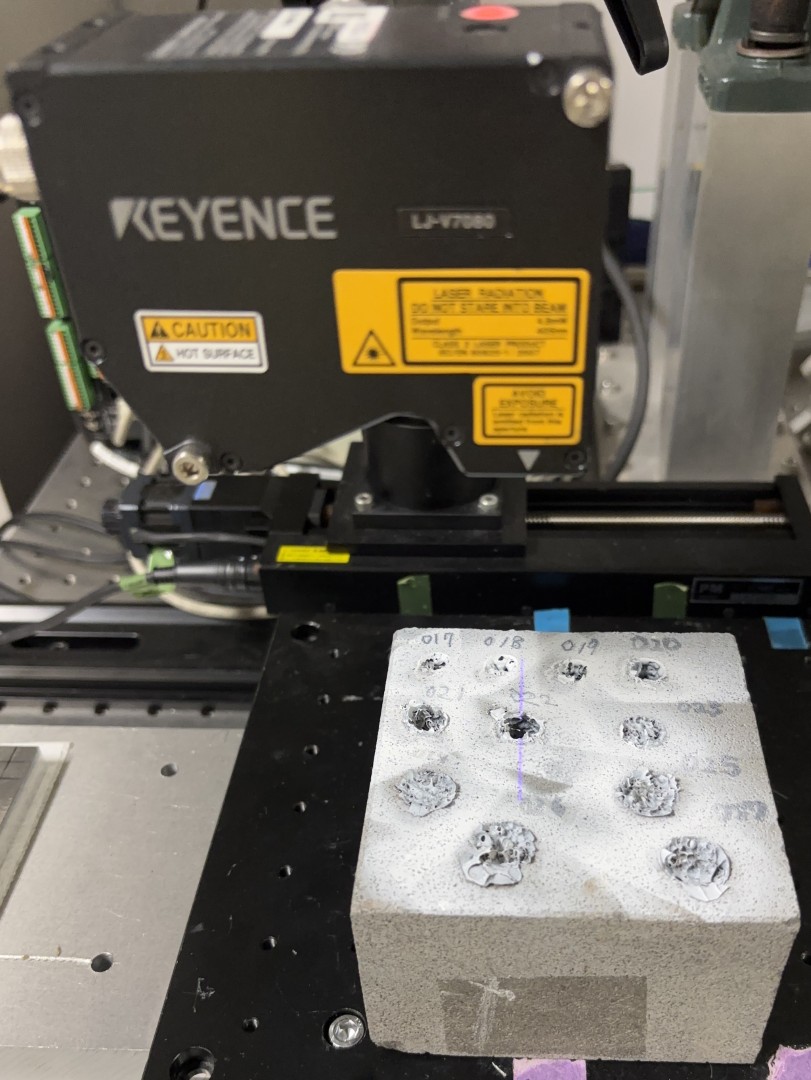}
            \label{fig:odm-photo}
        \end{minipage}
    \end{tabular}
    \caption{Optical displacement meter. (a) Schematic of the measurement system. (b)~Actual photo of the laser sensor and a sample.}
    \label{fig:odm-setup}
\end{figure}

After irradiation, the crater morphology formed on the sample surface was measured using an optical displacement meter (KEYENCE LJ-V7080) and an automatic one-dimensional motion stage (COMS) with a precision of 0.5~$\mu$m in height direction and 50~$\mu$m in horizontal direction. This measurement system is identical to that used in previous studies~\cite{Takizawa:2019,Omura:2020}. A schematic diagram and photograph of the measuring system are shown in Fig.~\ref{fig:odm-setup}.

After laser irradiation, dome-shaped bubbles are occasionally formed on the sample surface, sometimes covering the crater morphology. This hollow structure is clearly different from the molten wavy rim structures observed in microcraters. Therefore, these bubbles were removed prior to the crater-shape measurement.
To obtain stable surface measurement results by suppressing optical scattering, a thin white spray paint was applied to the surface before measurement.

\section{Experimental Results}
\label{sec:result}

\begin{figure}
  \centering
  {\setlength{\tabcolsep}{2pt}
    \begin{tabular}{@{}m{0.02\linewidth}cccc@{}}
                                                                                                          & $t=0~s$ & $t=0.1~s$ & $t=1~s$ & $t=5~s$ \\
      \raisebox{0.85cm}[0pt][0pt]{(a)}                        &
      \scalebar{\includegraphics[width=0.22\linewidth,clip,trim=490bp 160bp 610bp 240bp, scale=0.6]{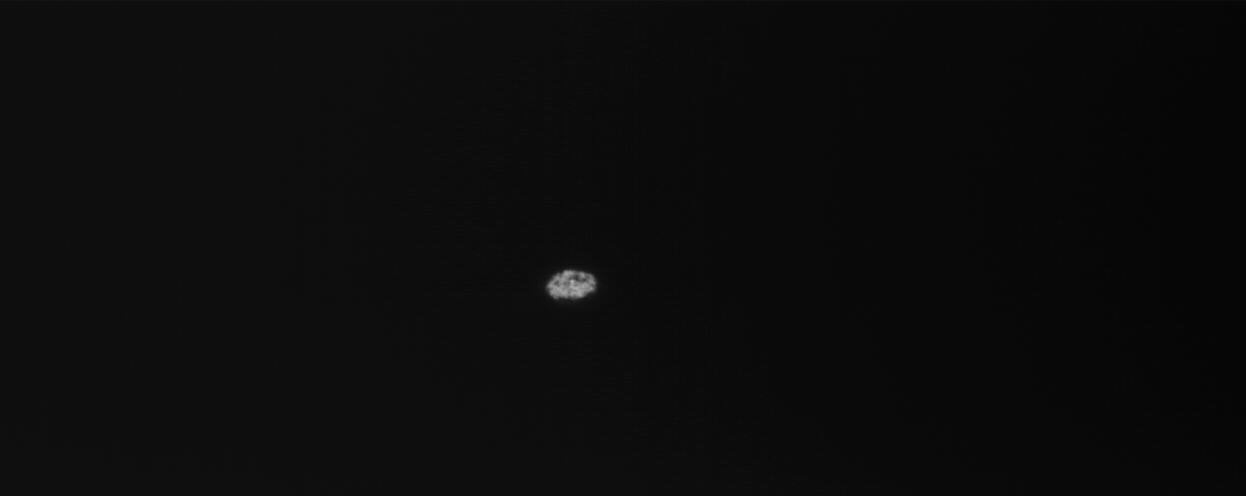}}{1246}{97}{5}{\milli\meter} &
      \includegraphics[width=0.22\linewidth,clip,trim=490bp 160bp 610bp 240bp, scale=0.6]{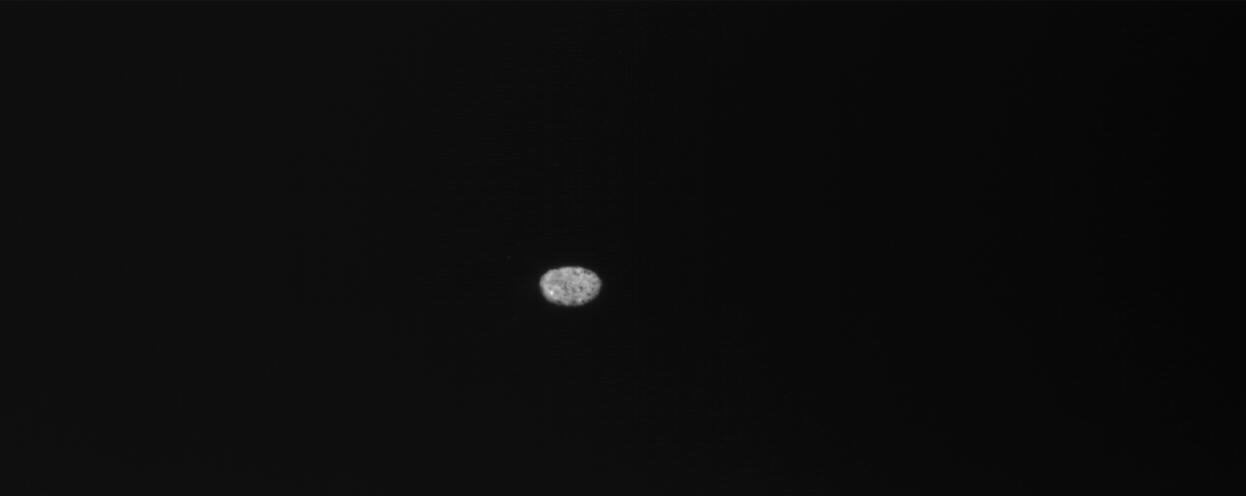} &
      \includegraphics[width=0.22\linewidth,clip,trim=490bp 160bp 610bp 240bp, scale=0.6]{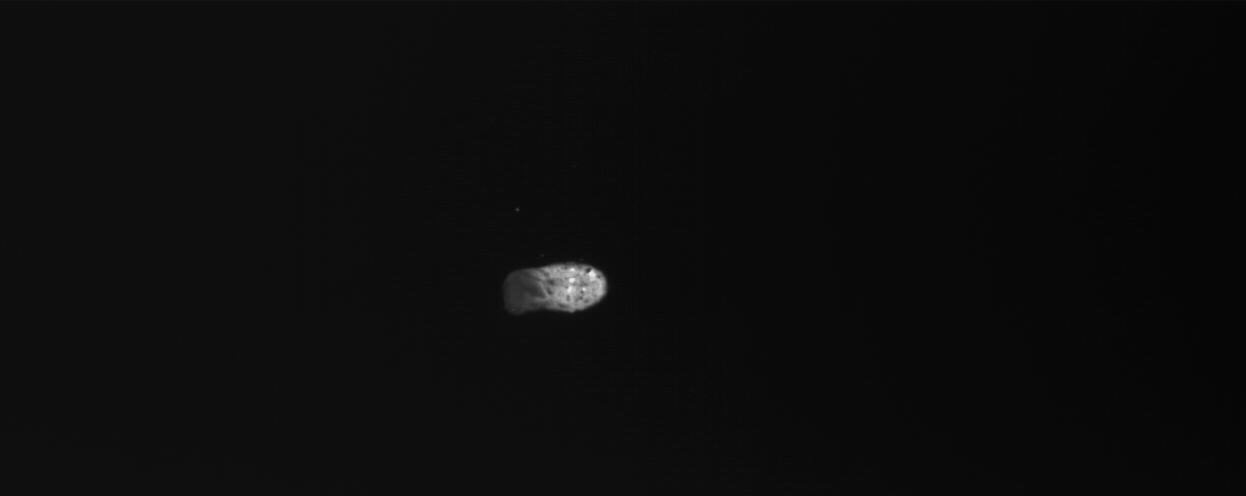}&
      \includegraphics[width=0.22\linewidth,clip,trim=490bp 160bp 610bp 240bp, scale=0.6]{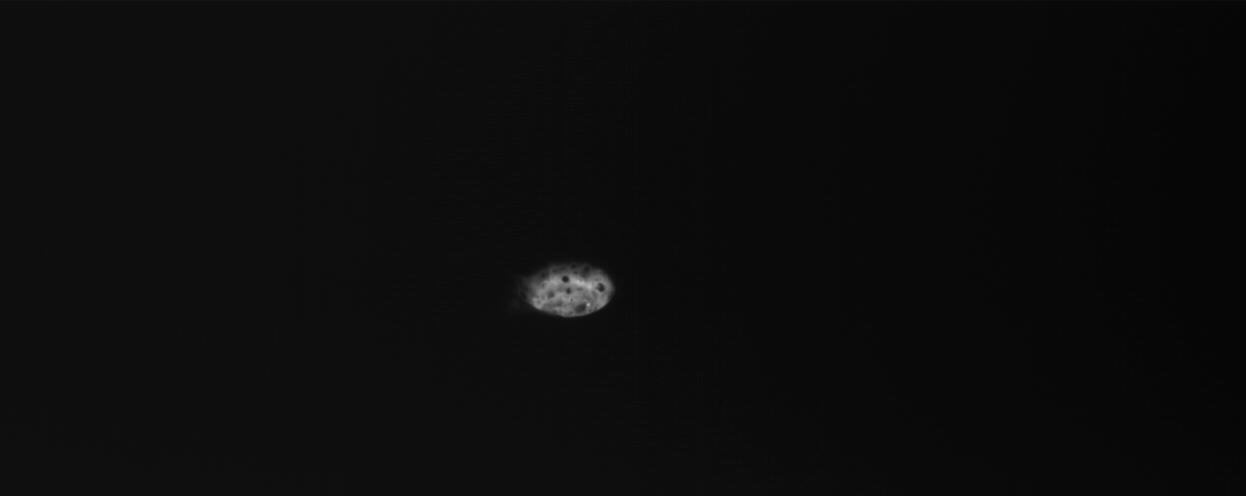} \\
      \raisebox{0.85cm}[0pt][0pt]{(b)}                       &
      \scalebar{\includegraphics[width=0.22\linewidth,clip,trim=410bp 150bp 600bp 180bp, scale=0.6]{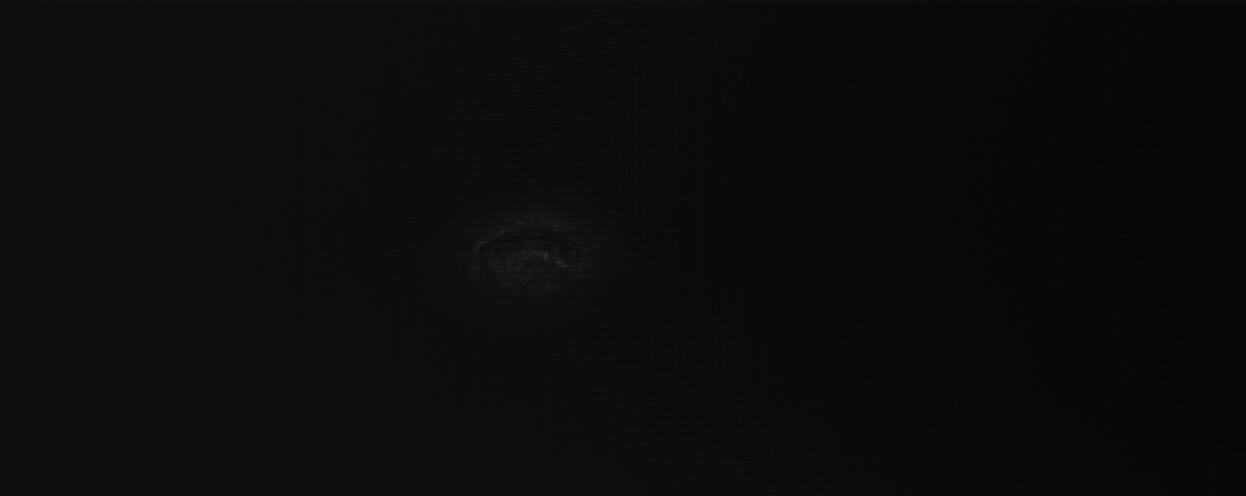}}{1246}{97}{5}{\milli\meter} &
      \includegraphics[width=0.22\linewidth,clip,trim=410bp 150bp 600bp 180bp, scale=0.6]{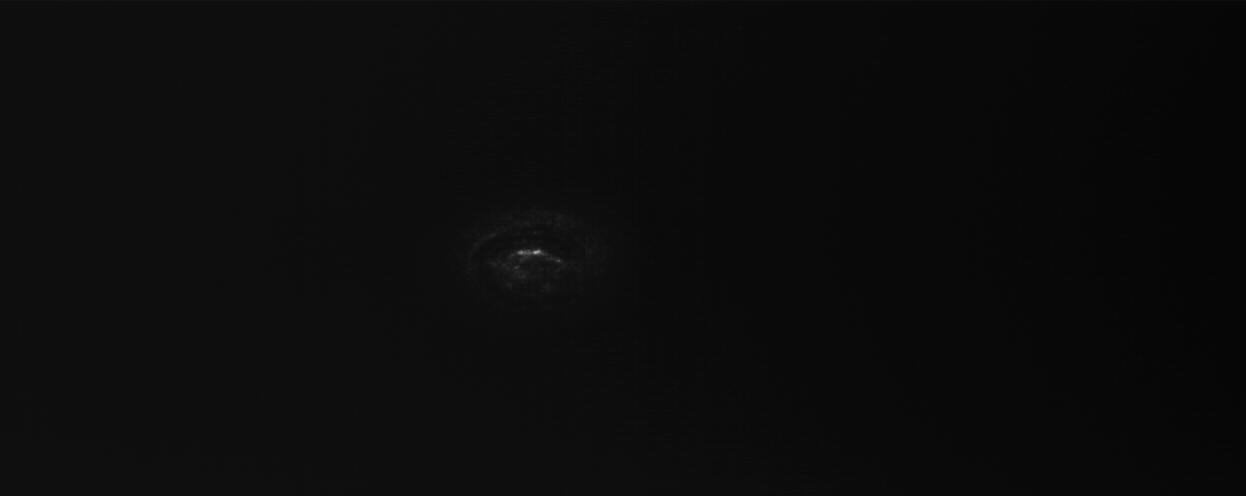} &
      \includegraphics[width=0.22\linewidth,clip,trim=410bp 150bp 600bp 180bp, scale=0.6]{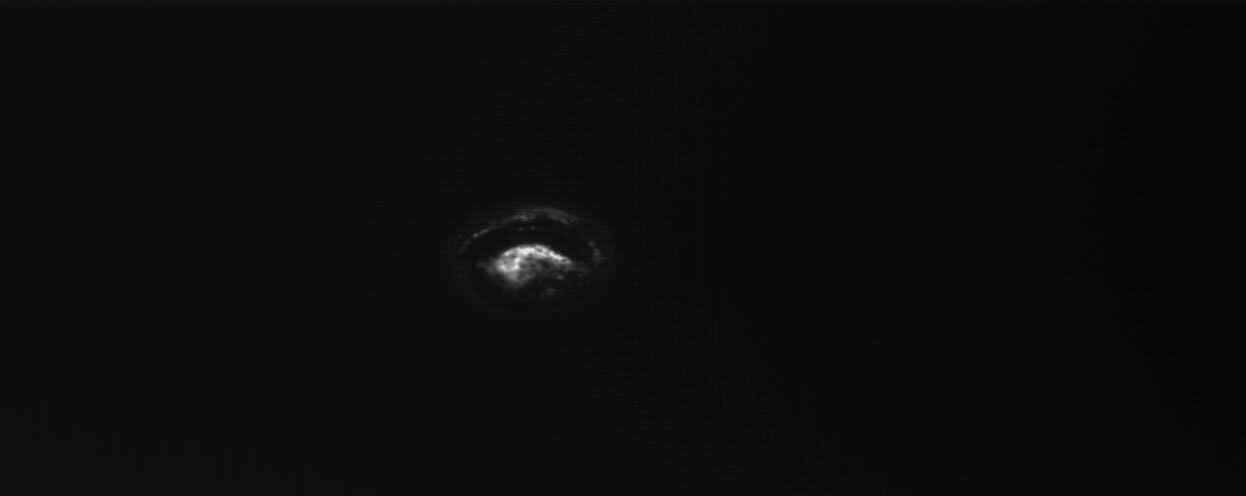} &
      \includegraphics[width=0.22\linewidth,clip,trim=410bp 150bp 600bp 180bp, scale=0.6]{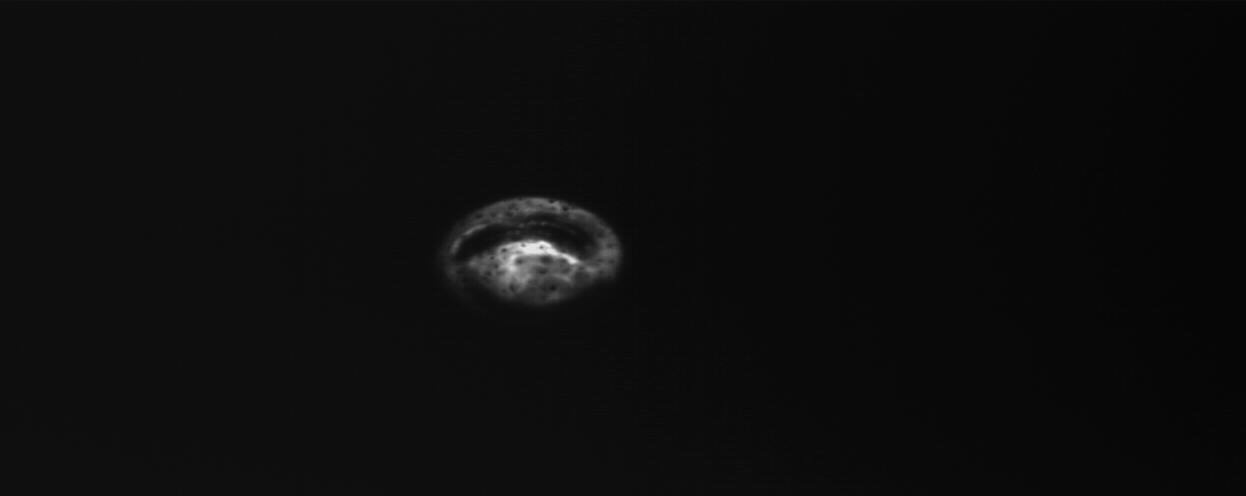} \\
      \raisebox{0.85cm}[0pt][0pt]{(c)}                         &
      \scalebar{\includegraphics[width=0.22\linewidth,clip,trim=640bp 180bp 460bp 220bp, scale=0.6]{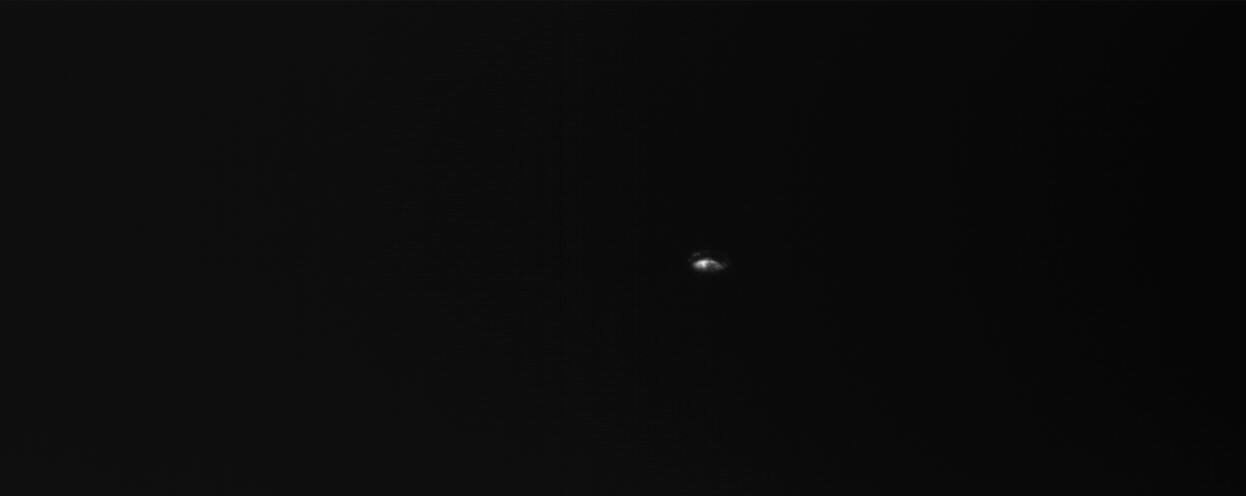}}{1246}{97}{5}{\milli\meter} &
      \includegraphics[width=0.22\linewidth,clip,trim=640bp 180bp 460bp 220bp, scale=0.6]{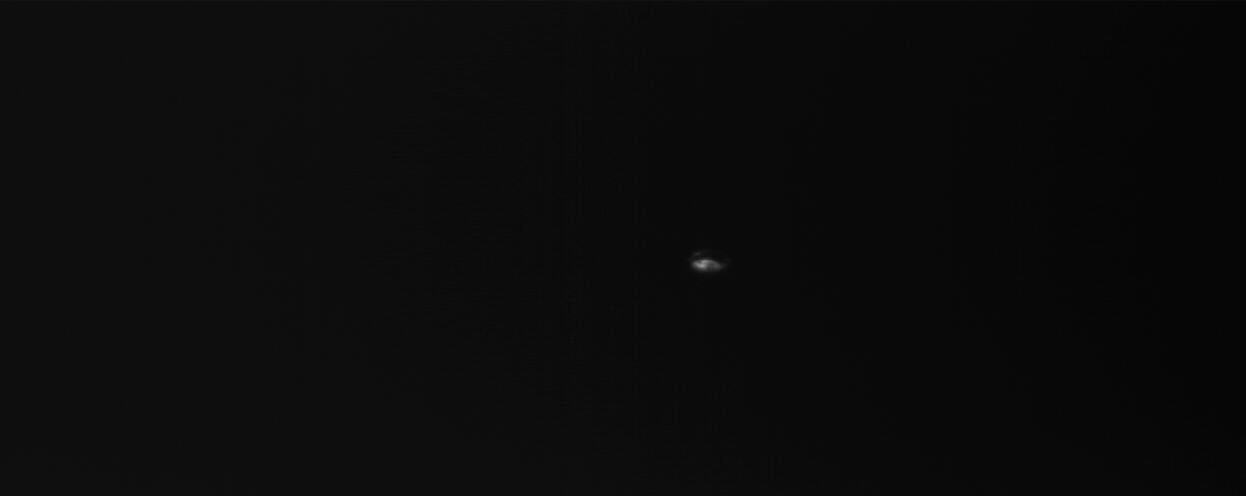}&
      \includegraphics[width=0.22\linewidth,clip,trim=640bp 180bp 460bp 220bp, scale=0.6]{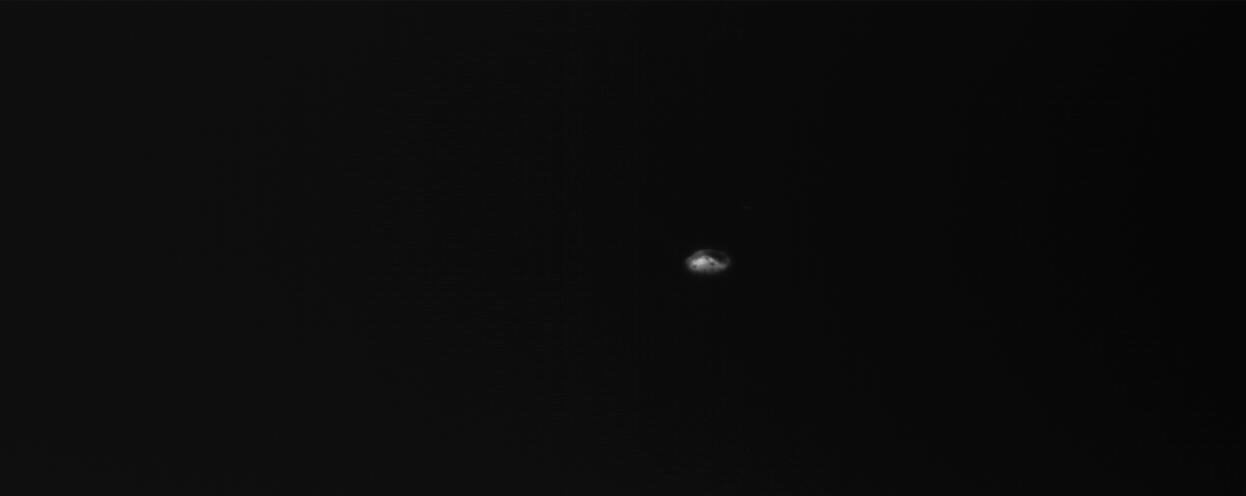} &
      \includegraphics[width=0.22\linewidth,clip,trim=640bp 180bp 460bp 220bp, scale=0.6]{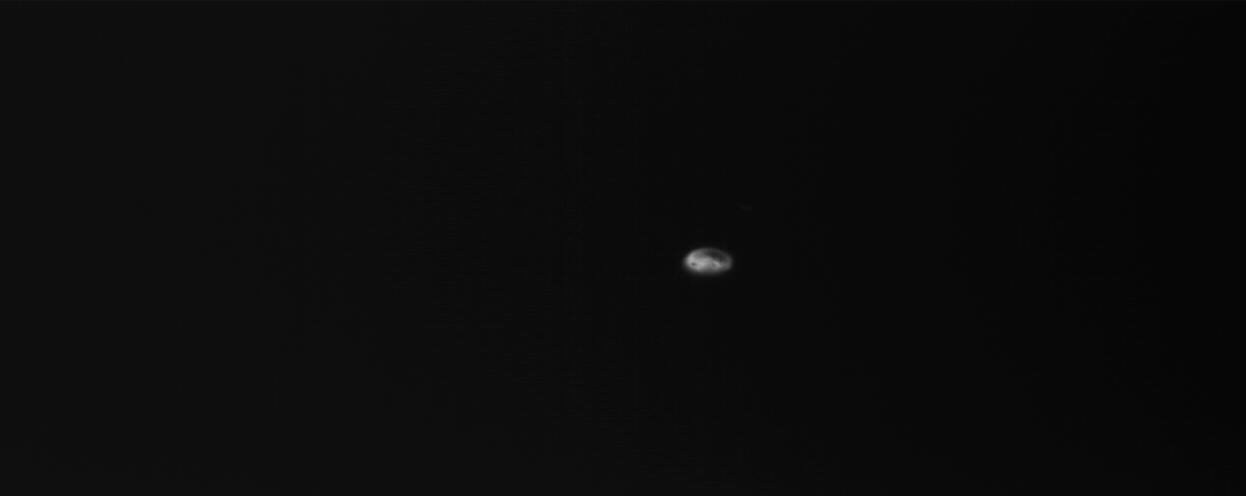} \\
    \end{tabular}
  }
  \caption{Snapshots of the crater formation processes during laser irradiation. Experimental conditions for each row: (a)~1000~W, DFS=0.1~m, (b)~1000~W, DFS=0.3~m, (c)~100~W, DFS=0.1~m. Vaporization due to the boiling can be confirmed. }
  \label{fig:result-snapshots}
\end{figure}

High-speed video data acquired by the high-speed camera were used to examine the temporal evolution of the sample surface during irradiation. Figure~\ref{fig:result-snapshots} shows representative snapshots of crater formation processes.
Images at time $t=0$, 0.1, 1, and 5~s are presented for different laser power or DFS. The reference time $t=$0~s was determined from the temporal variation of the total image luminance, corresponding to the onset of the laser irradiation.  
In the snapshots, progressive crater development accompanied by vigorous boiling is clearly observed as irradiation proceeds. 

\begin{figure}
  \centering
  {\setlength{\tabcolsep}{4pt} 
    \begin{tabular}{@{}ccc@{}} 
    & Photo & Profile \vspace{-0.7cm} \\ 
        \raisebox{2.0cm}[0pt][0pt]{(a)} & 
        \scalebar{\includegraphics[width=0.25\linewidth]{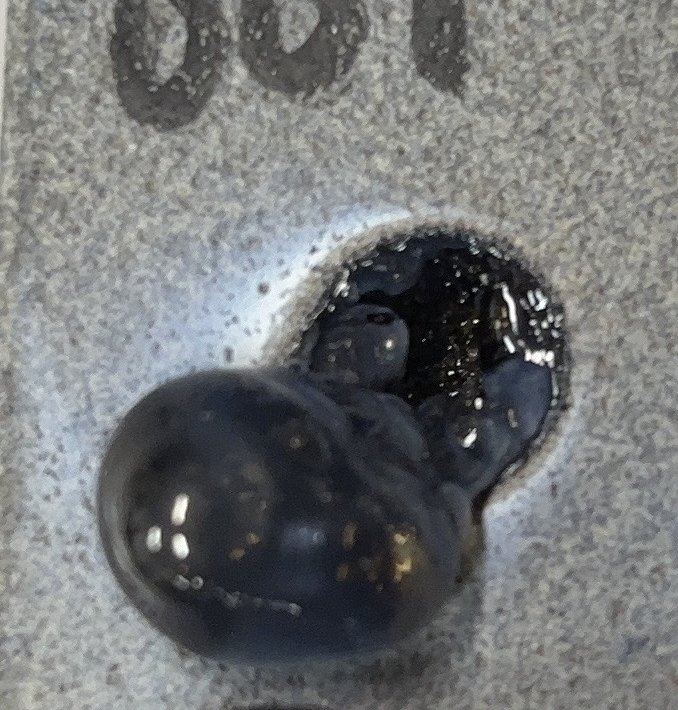}}{678}{30}{5}{\milli\meter} &  
        \includegraphics[width=0.55\linewidth]{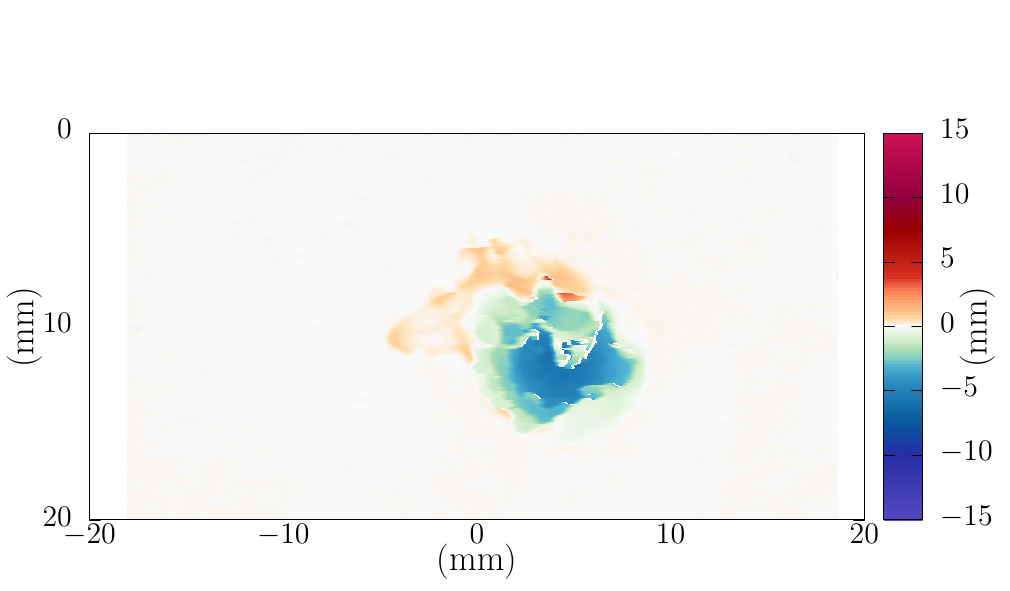}  \\
        
        \raisebox{2.0cm}[0pt][0pt]{(b)} & 
        \scalebar{\includegraphics[width=0.25\linewidth]{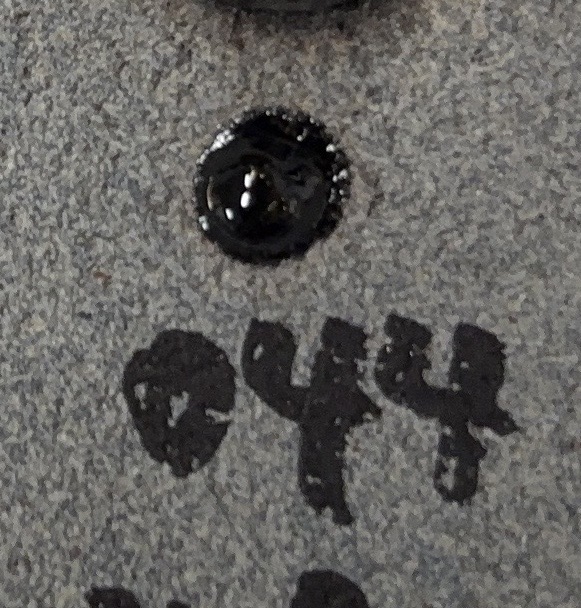}}{581}{35}{5}{\milli\meter} &
        \includegraphics[width=0.55\linewidth]{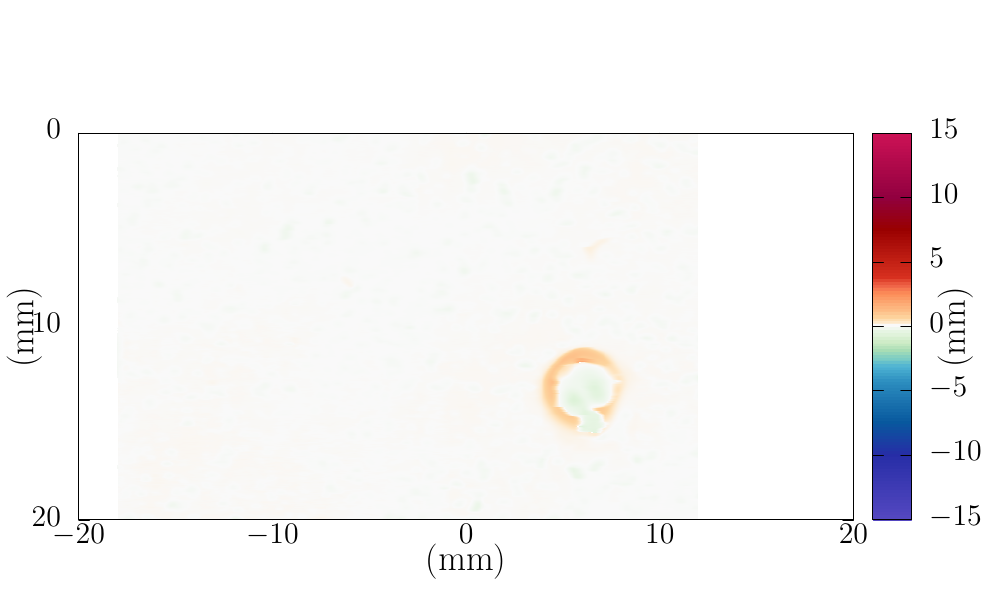}  \\
        
        \raisebox{2.0cm}[0pt][0pt]{(c)} & 
        \scalebar{\includegraphics[width=0.25\linewidth]{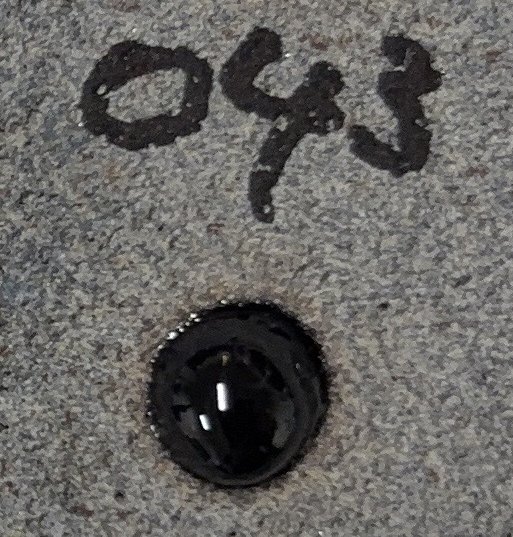}}{513}{30}{5}{\milli\meter}  &
        \includegraphics[width=0.55\linewidth]{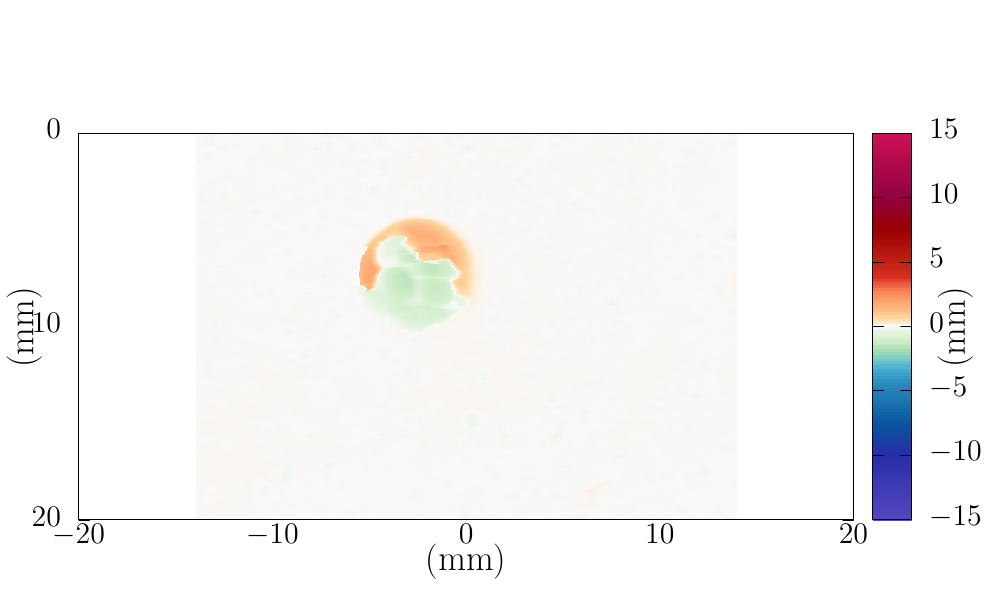}  \\
        
        \raisebox{2.0cm}[0pt][0pt]{(d)} & 
        \scalebar{\includegraphics[width=0.25\linewidth]{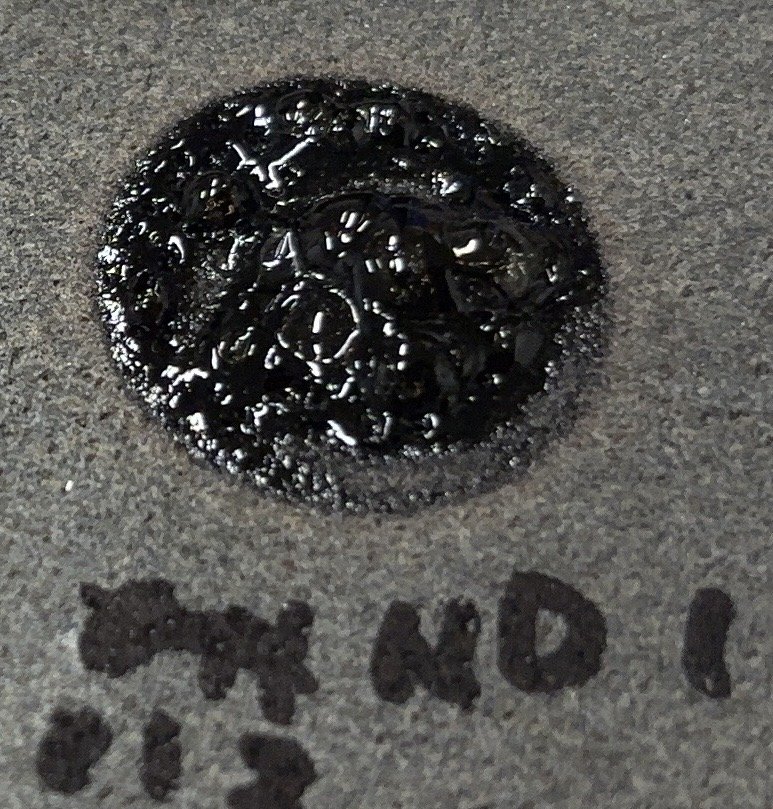}}{773}{30}{5}{\milli\meter}  &
        \includegraphics[width=0.55\linewidth]{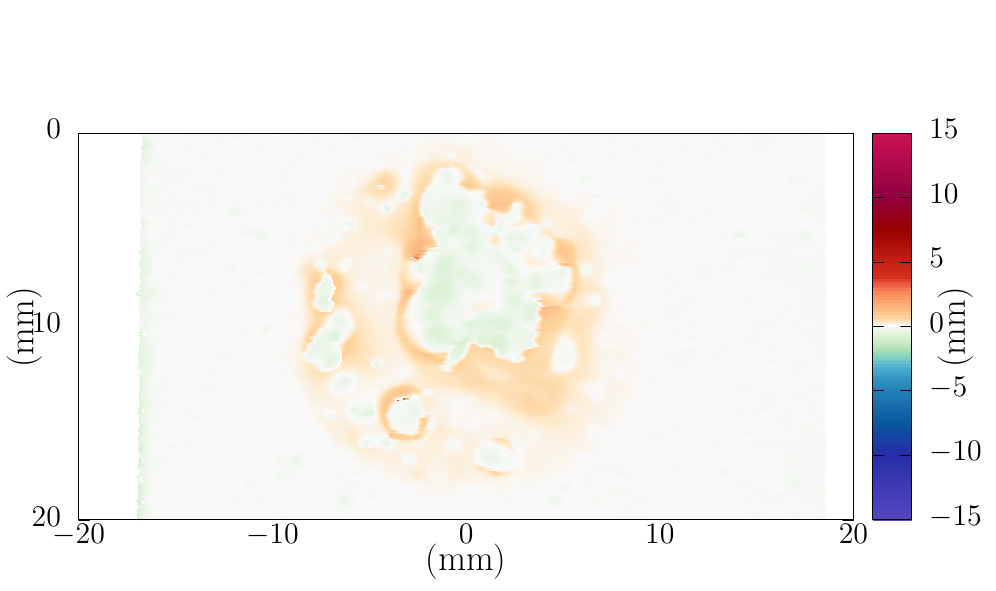} \\
    \end{tabular}
  }
  \caption{Crater shapes. Laser irradiation conditions for each row are: (a)~1000~W, 5~s, and DFS=0.1~m, (b)~100~W, 5~s, and DFS=0.1~m, (c)~100~W, 100~s, and DFS=0.1~m, and (d)~1000~W, 5~s, and DFS=0.3~m. Left and right columns indicate photos and corresponding measured height maps, respectively. White parts in the height maps corresponds to the initial base height or missing data points. Note that hollow bubbly parts are removed before surface-profile measuring.}
  \label{fig:result-photos}
\end{figure}

Photographs of the sample surface were taken after irradiation. Figure~\ref{fig:result-photos}~(left column) shows post-irradiation image examples. 
Figure~\ref{fig:result-photos} compares results obtained at 1000 or 100~W with DFS values of 0.1 or 0.3~m, and irradiation duration 5 or 100~s. The morphology of molten remnant structures depends on irradiation conditions. Larger average power and longer irradiation duration tend to produce larger melt structures. Larger DFS values result in wider but shallower structures.  Qualitatively, the observed results are natural. Although we are interested in the formation conditions of molten rim structures, it is difficult to directly measure the volume of molten (glassy) parts. As shown in Fig.~\ref{fig:result-photos}(a) left, hollow bubble formation frequently happens particularly when the irradiation energy is large. Besides, it is difficult to distinguish between molten and unmelted parts using the current measurement system. Therefore, we instead measure the crater cavity shape. To do that, bubbly parts are removed before measuring the surface profile shown in Fig.~\ref{fig:result-photos}~(right column).

Surface height profiles, obtained using the optical displacement meter after removing the bubbly sections, are presented in the right column of Figs.~\ref{fig:result-photos}. 
As can be seen, crater diameter and depth vary depending on irradiation conditions. Longer irradiation time and higher average power generally produce deeper craters. Conversely, increasing DFS results in shallower crater formation.

\section{Analysis and Discussion}
The total laser irradiation energy is defined as the product of the average power and irradiation duration,
\begin{equation}
  E = P_{\mathrm{ave}} t.
\label{eq:laser-energy}
\end{equation}
As shown in the previous section, crater morphologies formed by relatively long-duration laser irradiation exhibit a wide variety of forms. 
It is not straightforward to explain such diverse pattern formation using a single physical mechanism. As a first step approach, here we measure the most fundamental quantities, diameter and volume of the formed craters to investigate the physical processes governing crater formation.

\subsection{Energy Dependence of Crater Diameter and Volume}
\begin{figure}
    \centering
  \includegraphics[width=0.8\linewidth]{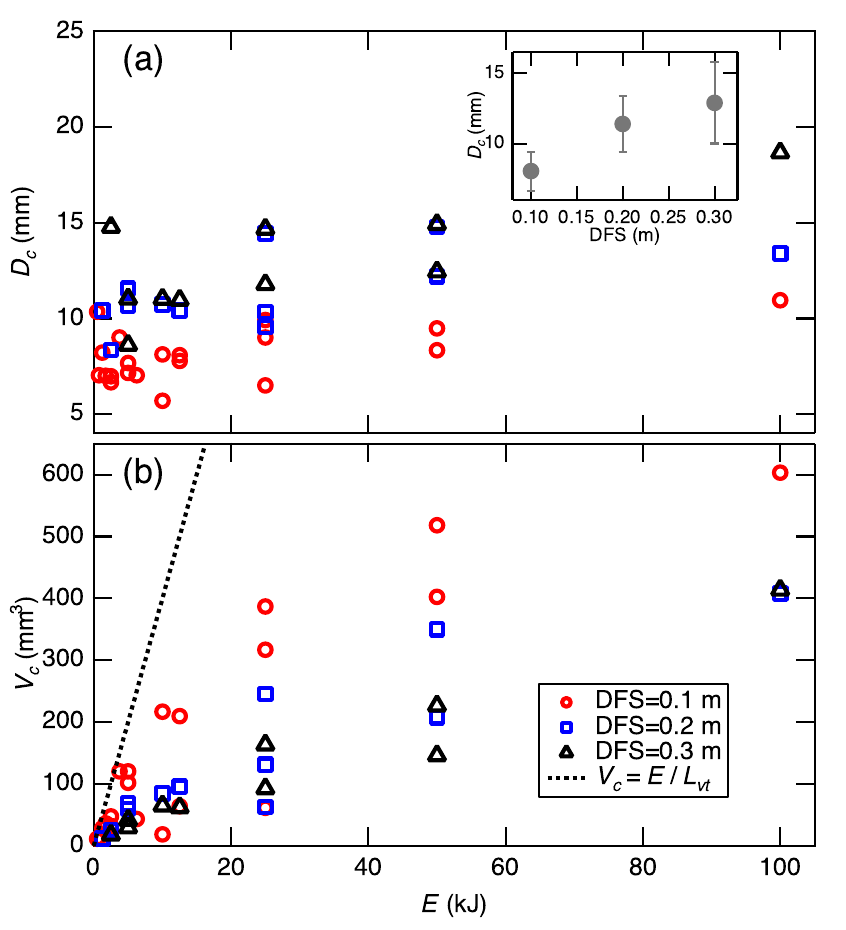}
  \caption{Relations between crater diameter $D_c$, volume $V_c$ and total irradiation energy $E=P_{\mathrm{ave}}t$. (a)~The relation $D_c$ vs. $E$ does not show clear data tendency. The $D_c$ rather correlates with DFS (inset). (b)~The relation $V_c$ vs. $E$ shows a clear positive correlation. The dashed line indicates an ideal relation, $V_c = E/L_{vt}$. As $E$ increases, indicating decreasing cratering efficiency because the deviation between black dashed line and data points grows. Each data point corresponds to an individual experimental run (no statistical averaging).}
  \label{fig:crater_diam_vol}
\end{figure}

Figure~\ref{fig:crater_diam_vol}(a) shows the relation between measured crater diameter $D_c$ and $E$. Here, $D_c$ is defined using the area of crater cavity $S_c$ as $D_c = \sqrt{4S_c/\pi}$. And, the crater cavity is defined as the depression relative to the initial surface level. Although a very weak positive correlation can be observed in Fig.~\ref{fig:crater_diam_vol}(a), a clear trend cannot be confirmed.  In this study, multiple experiments with the identical conditions have not been performed. Since the measurement precision is small enough, the measurement uncertainty will be smaller than the size of symbols in Fig.~\ref{fig:crater_diam_vol} (and also in Fig.~\ref{fig:crater_diam_vol}). 
This behavior contrasts with hypervelocity impact cratering, where crater diameter scales systematically with impact energy. The absence of such a correlation in laser irradiation experiments can be attributed to the fact that the irradiated area (equivalently diameter) is geometrically constrained by the laser spot size, which is set by the DFS rather than by the total deposited energy. In other words, the spatial extent of energy deposition, not its magnitude, primarily governs crater diameter in this experiment. Thus, $D_c$ shows a certain dependence on DFS (inset of Fig.~\ref{fig:crater_diam_vol}(a)). 
An increase in DFS enlarges the laser spot size, leading to larger crater diameters. 

The dependence of crater volume $V_c$ on total irradiation energy $E$ is shown in Fig.~\ref{fig:crater_diam_vol}(b). 
Unlike crater diameter, $V_c$ increases with increasing $E$, consistent with the energy consumption concept. The curved (convex upward) behavior seen in Fig.~\ref{fig:crater_diam_vol}(b) can be interpreted in terms of energy partition during laser irradiation. When a laser pulse is incident on the rock surface, the input energy is distributed into several processes, including heating, melting, vaporization, and thermal diffusion into the surrounding material. Only a fraction of the input energy contributes to the excavation (vaporization) that forms the crater.
If we denote the effective energy available for excavation as $E_e$, the crater volume $V_c$ can be related to $E_e$ through a scaling relation. In this framework, the observed relation between the input energy and crater volume reflects how the input energy is partitioned among the competing processes. The deviation from a simple linear relation therefore indicates that a substantial portion of the energy is dissipated typically by thermal diffusion rather than excavation. A similar energy partition is expected in micro-scale impact cratering, where only a fraction of the impact energy is converted into excavation (vaporization) energy.

During excavation, the target rock appears to melt and boil (Fig.~\ref{fig:result-snapshots}), suggesting that the supplied energy is consumed by heating, melting, and vaporization processes. 
The energy required to heat, melt, and vaporize unit-volume rock is denoted as $L_{vt}$. 
Based on reported values for granite and basalt, we roughly estimate the value of $L_{vt}$ for andesite as $L_{vt} = 0.025~\mbox{kJ/mm}^3$~\cite{Oglesby:2014}. 
The ideal (minimum) energy required to vaporize a crater volume $V_c$ is written as $E_V = L_{vt} V_c.$ This relation (by substituting $E_V=E$) is shown in Fig.~\ref{fig:crater_diam_vol}(b), as a black dashed line. 

As shown in Fig.~\ref{fig:crater_diam_vol}(b), $E$ and $E_V$ roughly agree in the low-energy regime. However, actual $V_c$ is much less than the volume expected by $E_V$ in large energy regime. 
This deviation suggests increasing energy loss during longer irradiation durations through heat conduction and other dissipative processes. 
Consequently, excavation efficiency decreases as the energy supply timescale increases. 
Larger DFS values also reduce excavation efficiency, consistent with reduced energy concentration due to increased spot size. 
Overall, efficient excavation is achieved when energy is supplied rapidly and concentrated within a small spatial region. 
This result is qualitatively natural because the widely spread long irradiation can cause dissipation such as thermal diffusion.  
To quantitatively analyze the result, we apply scaling method in the next section.

\subsection{Scaling of Crater Volume}

Following the dimensional-analysis approach commonly used in impact cratering studies~(e.g.~\cite{Holsapple:1982,Katsuragi:2016}), we introduce dimensionless parameters suitable for the current laser-irradiation situation.

Characteristic quantities should include $E$, $V_c$, $L_{vt}$, and material density $\rho$. 
In addition, as a characteristic length scale, thermal diffusion length is introduced,
\begin{equation}
\lambda_H = \sqrt{\kappa t},
\end{equation}
where $\kappa$ is thermal diffusivity of the target material and $t$ is irradiation time. We use a representative value of $\kappa=1\times10^{-6}$~m$^2$/s for rocks~\cite{Ji:2024}. To form molten rim structures, energy loss must be induced by thermal diffusion. Otherwise,  energy is efficiently consumed to vaporize the target material. We consider this energy loss is a key to understand the molten rim formation. Thus, we consider the thermal diffusion length here. Here we neglect other minor dissipation such as acoustic emission, radiation, and atmospheric heat transfer.

\begin{figure}
    \centering
  \includegraphics[width=0.8\linewidth]{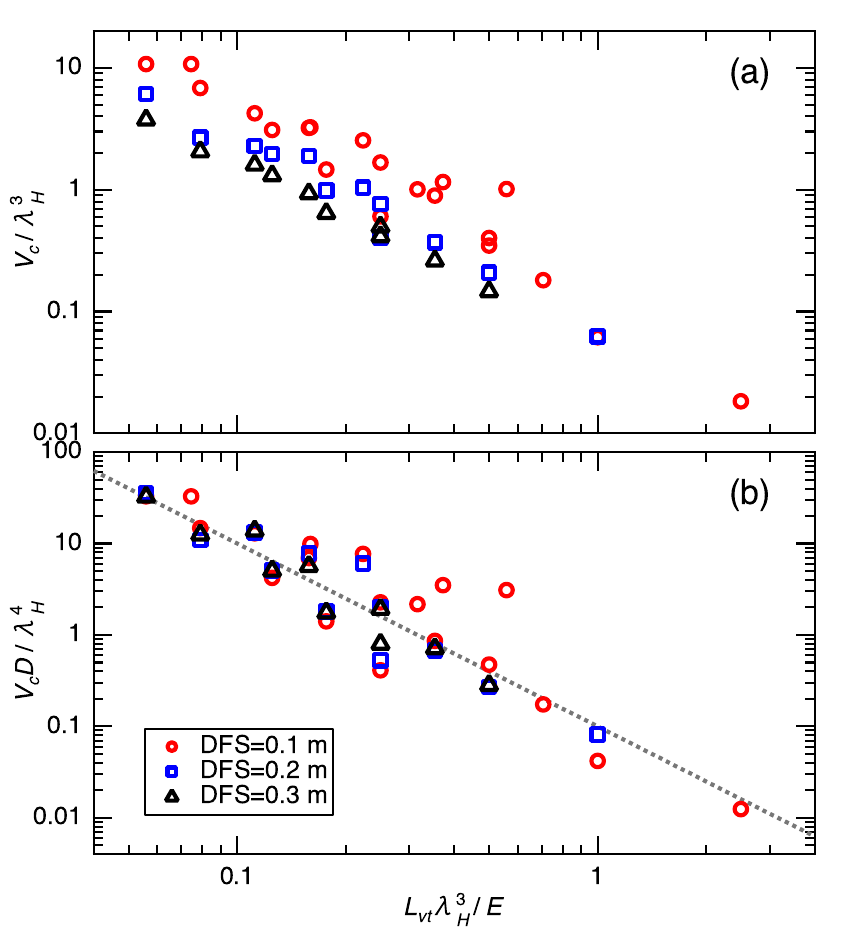}
  \caption{Dimensionless scaling of the crater volume. (a) Scaling of $V_c$ vs. $E$ using a characteristic length scale $\lambda_H = \sqrt{\kappa t}$ and $L_{vt}$. While all the data follow a similar scaling trend, systematic deviation by DFS can be confirmed. (b) Modified scaling by using a spot diameter $D$. All data are collapsed onto a simple scaling law represented by the broken line, $DV_c/\lambda_{H}^4 \sim (L_{vt}\lambda_{H}^3 /E)^{-2}$. Each data point corresponds to an individual experimental run (no statistical averaging).}
  \label{fig:Vc_E_scaling}
\end{figure}

By extending the conventional impact-cratering scaling in the strength regime, we obtain a scaling relation between $E$ and $V_c$:
\begin{equation}
\frac{V_c}{\lambda_H^3}
= C_0 \left( \lambda_H^3 \frac{L_{vt}}{E} \right)^{-\alpha_0},
\end{equation}
where $C_0$ and $\alpha_0$ are fitting dimensionless parameters.
To confirm the validity of this scaling,  $V_c/(\lambda_H^3)$ is plotted as a function of $\lambda_H^3 L_{vt}/E$ in Fig.~\ref{fig:Vc_E_scaling}(a). One can confirm the parallel scaling tendency for all data.   However, the coefficient $C_0$ depends on DFS. This implies the incomplete scaling.

The residual DFS-dependence in Fig.~\ref{fig:Vc_E_scaling}(a) suggests that the spatial concentration of energy plays an independent role in governing crater growth, beyond what is captured by the thermal diffusion length alone. Physically, the laser spot diameter $D$ indicates the lateral scale over which energy is deposited, while $\lambda_H = \sqrt{\kappa t}$ characterizes how far heat diffuses during irradiation time. The ratio $D/\lambda_H$ represents the competition between energy concentration and thermal diffusion. 
Incorporating this additional dimensionless factor, we propose the improved scaling relation,
\begin{equation}
\frac{D}{\lambda_H} \frac{V_c}{\lambda_H^3}
= C \left( \lambda_H^3 \frac{L_{vt}}{E} \right)^{-\alpha},
  \label{eq:final_scaling}
\end{equation}
Here, $C$ is a dimensionless coefficient and $\alpha$ is a dimensionless scaling exponent, both determined by fitting.
The spot diameter is estimated as $D=0.064\times$DFS, based on the beam divergence geometry of the light source system, validated by high-speed video observations. 

The corresponding normalized data are plotted in Fig.~\ref{fig:Vc_E_scaling}(b). As shown in Fig.~\ref{fig:Vc_E_scaling}(b), the quality of data collapse is significantly improved. From the data fitting, we obtain $\alpha=2$ and $C =1\times 10^{-1}$. This exponent differs significantly from the classical impact-cratering scaling in the strength regime, where $\alpha=1$ is expected when crater volume scales linearly with the ratio of input energy to mechanical strength. The difference can be qualitatively understood as follows. In the strength-dominated regime, the energy consumption linearly relates to the crater volume, giving $\alpha=1$. In the present laser-irradiation case, however, a competing energy sink arises from thermal diffusion into the surrounding material. The energy lost to diffusion scales with the diffusion volume $\sim \lambda_H^2 D$.
As a result, the fraction of input energy available for excavation decreases. This trend is directly visible as the convex-upward curvature in the $V_c$ - $E$ relation shown in Fig.~\ref{fig:crater_diam_vol}(b). When this time-dependent diffusion loss is incorporated into the dimensionless scaling, the linear ($\alpha=1$) dependence is modified to $\alpha=2$. This $\alpha$ value qualitatively suggests that the energy efficiency for cavity formation is significantly reduced by long irradiation. 
We note, however, that this interpretation is heuristic, and a rigorous first-principles derivation of this exponent remains a subject for future work.
This contrasts with the classical strength-dominated scaling ($\alpha=1$).
In the present case, the effective energy available for excavation decreases with increasing diffusion volume, leading to a nonlinear scaling. 
This highlights that energy dissipation, rather than strength, controls the scaling in the present regime.

\begin{figure}
    \begin{tabular}{cc}
        \begin{minipage}{0.47\hsize}
            \centering
            \begin{tikzpicture}
              \draw[red, thick] (-0.5, 2) -- (1, -1);
              \draw[red, thick] (0.5, 2) -- (-1, -1);
              \node at (1.1, 1) {Laser beam};
              \node at (-2, 2) {(a)};
              \draw[thick] (-2, -1) -- (2, -1);
              \node at (0, -1.3) {Sample};
              \draw[<->] (-1, -0.95) -- (1, -0.95);
              \node at (0, -0.7) {$D$};
            \end{tikzpicture}
        \end{minipage} &
        \begin{minipage}{0.47\hsize}
            \centering
            \begin{tikzpicture}
              \draw (0,1) circle[radius=1];
              \draw[<->] (-1,1) -- (1,1);
              \node at (0, 0.6) {$D_i$};
              \node at (2, 1) {Impactor};
              \node at (-2,2) {(b)};
              \draw[thick,->,dashed] (0,0) -- (0,-1.5);
              \node at (0.5, -0.6) {$v$};
              \draw[thick] (-2, -1.5) -- (2, -1.5);
              \node at (0, -1.8) {Sample};
            \end{tikzpicture}
        \end{minipage}
    \end{tabular}
    \caption{Correspondence between (a)~laser and (b)~impact. The spot size roughly corresponds to the impactor size. The impact time scale is estimated by $D_i/v$. The total energy in the impact case is the impact kinetic energy.}
    \label{fig:laser-impact}
\end{figure}
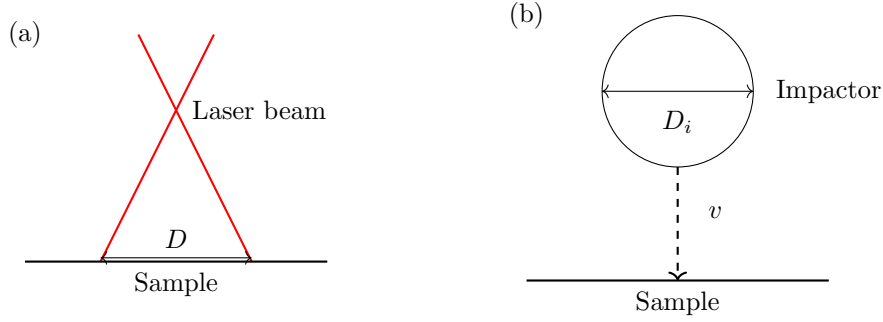

Here, we define excavation efficiency $\eta$ as
\begin{equation}
\eta = \frac{L_{vt} V_c}{E}.
  \label{eq:eta}
\end{equation}
Perfect efficiency meaning all energy is consumed by vaporization corresponds to $\eta = 1$.
Using the experimentally derived scaling relation of Eq.~(\ref{eq:final_scaling}), $\eta$ can be written as:
\begin{equation}
\eta = \frac{C}{\lambda_H^2 D}\frac{E}{L_{vt}}.
\end{equation}
The smaller $\eta$ indicates that energy loss due to thermal diffusion and melting (without vaporization) is significant. In such situation, molten rim structure can be clearly observed. Due to the bubble formation in the molten rim, it is difficult to directly measure the molten volume in this study. However, by the qualitative inspection, we confirm that large $\eta$, typically obtained for shorter irradiation durations,  results in less molten rim formation. Therefore, we consider this efficiency $\eta$ can be used for the indicator of melt formation.  
In our experiments, molten rim structures were observed under almost all irradiation conditions except those with very small total energy $E$, where insufficient heating prevented rim formation. This observation is qualitatively consistent with the $\eta$ framework: at very small $E$, $\eta$ approaches unity, meaning nearly all input energy is consumed by vaporization with little residual heat available for melting. At intermediate to large $E$, $\eta$ decreases and molten rims become prominent. We note that hollow dome-shaped bubbles occasionally form alongside the molten rim, particularly at large $E$, complicating direct quantification of the melt volume. Nevertheless, the overall trend supports the use of $\eta<1$ as a qualitative indicator for molten rim formation.

\subsection{Application to impact-induced microcrater formation}
Before applying the laser-derived scaling to impact cratering, we address the physical basis for this analogy. Although laser irradiation and impact cratering differ fundamentally in their energy deposition timescales and mechanisms, we argue that the final crater morphology is governed in both cases by the energy partitioning efficiency $\eta$. In the strength-dominated regime, crater volume is expected to scale linearly with input energy ($V_c \sim E$, i.e., $\alpha=1$). The nonlinear trend observed in Fig.~\ref{fig:crater_diam_vol}(b), however, reveals that excavation efficiency decreases as irradiation duration increases, due to growing energy loss via thermal diffusion. This competition between excavation and diffusion loss is captured by the scaling exponent $\alpha=2$ obtained in our experiments. The key insight is that this competition is governed not by the mode of energy deposition, but by the ratio of deposited energy to the heat conducted away during the characteristic timescale of the process. Molten rim formation thus occurs when a sufficient fraction of the deposited energy remains as heat within the target rather than being consumed by vaporization. Then, the condition can be expressed through $\eta$ to first order, regardless of the energy source: laser irradiation or impact. 
This supports the application of the laser-derived scaling to impact-induced microcrater formation. We note, however, that the values of $C$ and $\alpha$ 
may differ between the two cases, and direct impact experiments are needed for quantitative validation. 

Based on this analogy, we translate the experimental parameters into impact conditions. The projectile kinetic energy $E_K \sim \rho D_i^3 v^2$ s taken to correspond to the total laser energy $E$. Here, numerical prefactors are omitted to focus on the order of magnitude estimate. 
In this expression, $D_i$ and $v$ are the diameter of the impactor and the impact velocity, respectively. Then, the impact timescale $t_\mathrm{imp}$ is defined as: 
$t_\mathrm{imp} = D_i/v.$ 
By using these relations, we obtain,
\begin{equation}
\eta = C \frac{\rho D_i v^3}{\kappa L_{vt}}.
  \label{eq:eta_impact}
\end{equation}
The excavation efficiency $\eta$ defined in Eq.~(\ref{eq:eta}) measures the fraction of input energy converted into vaporization-driven material removal. When $\eta \simeq 1$, nearly all deposited energy drives vaporization, leaving little residual thermal energy to sustain a melt layer at the crater rim. Conversely, when $\eta \ll 1$, a substantial fraction of energy is dissipated by thermal diffusion into the surrounding material, potentially sustaining a molten zone at the crater periphery. We therefore propose that $\eta < 1$ constitutes a necessary condition for molten rim formation. (A more complete criterion would require that the local temperature exceeds the melting point over a finite volume.) If all energy is consumed by vaporization, no melt can persist at the rim. However, we stress that $\eta < 1$ is not a sufficient condition. Even when $\eta < 1$, the residual energy may be carried away by thermal conduction without producing a visible melt rim. A complete criterion for molten rim formation would additionally require that the energy deposited within the thermal diffusion volume exceeds the melting enthalpy of the target material over a sufficient volume. We are not able to quantify this condition precisely only from the present experiments. Therefore, we use $\eta < 1$ as a necessary (but not sufficient) condition, and estimate the upper bound on impact velocity for molten rim formation. 

To estimate the typical order of magnitude of the impact velocity, we substitute representative values (by neglecting prefactors), $\rho\simeq 10^3$~kg/m$^3$, $L_{vt}\simeq 10^{10}$~J/m$^3$ \cite{Oglesby:2014}, $\kappa\simeq 10^{-6}$~m$^2$/s \cite{Ji:2024}, and $D_i=10^{-6}$~m. Note that the prefactors of these quantities are completely neglected for this order estimate. This assumption is unlikely to affect the order-of-magnitude estimate significantly. For instance, even replacing $\rho \simeq 10^3$~kg/m$^3$ with the actual rock density $\rho \simeq 2.5 \times 10^3$~kg/m$^3$ would modify the velocity estimate by only a factor of $\sim 1.4$, which is within the order-of-magnitude accuracy of this analysis. The estimate is robust to order-of-magnitude variations in parameters.

Then, $\eta<1$ results in $v_\lesssim 5\times10^2$~m/s. Even if these parameter values vary within one order of magnitude, the estimated impact velocity remains within the same order of magnitude ($\sim 10^2$–$10^3$~m/s) because of $v^3$ dependence in Eq.~(\ref{eq:eta_impact}). This velocity range is consistent with the idea of secondary impact producing microcraters with molten rims~\cite{Harries:2016,Matsumoto:2018,Gu:2025}. Furthermore, this impact velocity is consistent with the estimate to form wavy rim structure based on hydrodynamic instability. To explain the peculiar protrusions observed on the microcrater rims, $v \simeq 10^2$~m/s is computed based on the Rayleigh-Taylor instability~\cite{Katsuragi:2015}. By combining all these estimates, we consider that the relatively low-speed ($v \simeq 10^2$~m/s) secondary impacts produce microcraters. Then, the molten and wavy rim structures can be explained by the energy loss and hydrodynamic instability. 

It is worth noting that the successful application of this scaling across very different spatial ($\mu$m vs. cm) and temporal ($\mu$s vs. s) scales is enabled precisely by the dimensionless formulation. The competition between $\lambda_H = \sqrt{\kappa t}$ and $D$ captures the same physical balance regardless of the absolute scale. Both cases are likely governed by similar dimensionless numbers. 

There are some limitations and constraints associated with this scaling relation. (i)~Obviously, the current scaling relation is relevant only when $\eta \leq 1$.  (ii)~For large impact cratering, gravity dominates the cratering. Thus, a wavy molten rim cannot be formed. The current discussion can be applied only for the small-scale cratering which should not be much larger than the capillary length scale. (iii)~Cooling rate is also crucial to create molten rim structure, although we have not discussed this effect in this paper. In small-scale phenomena, surface-volume ratio becomes large, resulting in rapid cooling. Intuitively, one might associate molten rims with high-energy impacts. However, low-speed impact is rather necessary to allow thermal diffusion and rapid cooling. Moreover, (iv)~we have not discussed the lower limit of $v$. It is impossible to evaluate it from the current experimental results. To form molten rims, an impact at a moderate speed (neither too slow nor too fast) is necessary. These factors must be properly considered to precisely evaluate actual microcrater structures. These factors remain to be explored in future studies.

From the viewpoint of laser material processing, the obtained scaling also offers practical guidance for minimizing energy loss during cutting and drilling operations. For the application to impact cratering, we have implicitly assumed that $C$ and $\alpha$ are independent of the energy deposition mechanism. In addition, the shockwave propagation and high-pressure conditions are not considered.  Such details are neglected in this study, for the sake of simplicity. This assumption is reasonable at the order-of-magnitude level. However, direct hypervelocity impact experiments would be needed to validate it quantitatively. While shock compression is not explicitly considered here, its primary role is to localize energy deposition, which is effectively parameterized by $\eta$ in the present framework. The present scaling therefore captures the post-deposition energy redistribution rather than the detailed shock physics. Furthermore, in the regime of high-intensity irradiation, the impulse delivered by the recoil of the evaporating mass becomes the dominant factor in crater formation, providing a physical bridge to the shock-induced deformation in hypervelocity impact events.

\section{Conclusion}
We have presented systematic laser-induced cratering experiments as a laboratory analogue for molten-rim microcrater formation on planetary surfaces. 
By systematically varying the laser power and irradiation duration, craters with molten rim structures were clearly observed. 
We modified the conventional crater scaling to account for energy loss through thermal diffusion, and derived a novel scaling relation for laser-induced excavation.
In our experimental conditions, a portion of the laser energy is consumed to vaporize target material. Therefore, the detailed analysis of crater volume provides the vaporization efficiency scaling. To obtain the relevant scaling, we have to consider thermal diffusion length scale. From the obtained scaling form and the typical material parameter values, we estimated typical impact velocity to form microcraters with molten rim structures. Owing to the scaling form, our macroscopic and long-duration laser irradiation experiment could mimic the competition between excavation (vaporization) and thermal diffusion during the energy injection timescale. By applying the scaling result to actual planetary impact situation, we find that the typical impact velocity for forming $1$~$\mu$m microcraters with molten rims should be less than  $5 \times 10^2$~m/s. In addition, to form wavy rim structures, the impact velocity should be in the order of $10^2$~m/s. The estimated values are consistent with the recent consideration for the origin of microcraters that were presumably produced by secondary impact~\cite{Harries:2016,Matsumoto:2018,Gu:2025}.

\section*{Conflict of Interest disclosure}
The authors declare there are no conflicts of interest for this manuscript.

\acknowledgments
This work was partially supported by JSPS KAKENHI Grant Number JP24H00196 and JST ERATO Grant Number JPMJER2401.

\bibliography{microcrater}

\end{document}